\documentclass[12pt,letterpaper]{report}

\usepackage{natbib}
\usepackage[french]{babel}
\usepackage[latin1]{inputenc}
\usepackage{amsmath,amssymb}
\usepackage{graphicx}
\usepackage{epsfig}
\usepackage{xspace}
\usepackage{layout}
\usepackage{setspace}
\usepackage{subfigure}

\usepackage{hangcaption}

\newcommand{\titre}{La recherche de naines brunes et d'exoplan\`{e}tes:\\ d\'{e}veloppement d'une technique d'imagerie multibande}
\newcommand{\nom}{Christian Marois}
\newcommand{\mois}{Juillet}  
\newcommand{\annee}{2004}

\begin{document}

\setlength{\parskip}{0.2cm}
\setlength{\oddsidemargin}{3cm}
\setlength{\evensidemargin}{1.7cm}
\pagestyle{headings}
\singlespacing

\pagenumbering{roman}
\thispagestyle{empty}
\large
\begin{center}
Université de Montréal
\end{center}

\vspace{1.5cm}
\begin{center} 
 \titre
\end{center}

\vspace{1.5cm}
\begin{center} 
 par\\ \vspace{0.5cm}
 \nom\\\vspace{0.5cm}
 Département de Physique\\
 Faculté des arts et des sciences
\end{center}

\vspace{1.5cm}
\begin{center}
 Th\`{e}se pr\'{e}sent\'{e}e \`{a} la Facult\'{e} des \'{e}tudes sup\'{e}rieures\\
en vue de l'obtention du grade de\\
Philosophi\ae \hspace{0.1cm}Doctor (Ph.D.)\\
en physique\\
\end{center}

\vspace{1.5cm}
\begin{center}
 \mois , \annee
\end{center}

\vspace{1.5cm}
\begin{center}
 \copyright \nom , \annee
\end{center}
\normalsize
\pagebreak[4]

\pagebreak[4]

\clearpage

\large
\setlength{\topskip}{1cm}
\begin{center}
Université de Montréal\\
Faculté des études supérieures
\end{center}

\vspace{1cm}

\begin{center}
Cette th\`{e}se intitulée:
\end{center}

\vspace{0.5cm}
\begin{center}
\titre
\end{center}
\vspace{1.5cm}
\begin{center}
présentée par:
\end{center}
\begin{center}
\nom
\end{center}
\vspace{1.5cm}
\begin{center}
a \'{e}t\'{e} \'{e}valu\'{e}e par un jury compos\'{e} des personnes suivantes:
\end{center}
\vspace{0.5cm}
\begin{tabbing}
\hspace{2.5cm}Pierre Bastien, \= \hspace{0.8cm}président du jury\\
\hspace{2.5cm}Daniel Nadeau, \> \hspace{0.8cm}directeur de recherche\\
\hspace{2.5cm}René Doyon, \> \hspace{0.8cm}codirecteur de recherche\\
\hspace{2.5cm}Tony Moffat, \> \hspace{0.8cm}membre du jury\\
\hspace{2.5cm}Jean-Luc Beuzit, \> \hspace{0.8cm}examinateur externe\\
\end{tabbing}

\vspace{1.cm}
\begin{tabbing}
\hspace{2.5cm}Thèse acceptée le: 8 octobre 2004\hspace{5cm}
\end{tabbing}
\normalsize
\onehalfspacing
\pagebreak[4]

\pagebreak[4]
\chapter*{Sommaire}
\addcontentsline{toc}{chapter}{Sommaire}
\thispagestyle{myheadings}
\markboth{}{}
Le projet se résume à faire l'observation, dans l'infrarouge, de plusieurs étoiles proches afin de détecter par imagerie directe des naines brunes et des exoplan\`{e}tes. La technique utilisée est celle de l'imagerie différentielle simultanée multibande. Cette technique consiste à obtenir des images simultanées à différentes longueurs d'onde afin de les combiner pour atténuer les tavelures atmosphériques et les structures provenant des erreurs de front d'onde statiques et quasi-statiques dans le but d'atteindre la limite imposée par le bruit de photon. Les longueurs d'onde sont sélectionnées près de la bande d'absorption du méthane à 1,6~$\mu $m de sorte qu'un objet méthanique émet plus à une longueur d'onde qu'aux autres. Ce choix permet de conserver une grande fraction du signal d'un compagnon méthanique à toutes les séparations angulaires lors de la combinaison des images. Un modèle analytique de la fonction d'étalement appuyé par des simulations numériques est d'abord présenté afin d'estimer les performances. Une caméra infrarouge à trois longueurs d'onde (TRIDENT) mise au point pour mettre en \oe uvre cette technique d'observation est par la suite décrite. Les résulats d'un sondage de 35 étoiles effectué au télescope Canada-France-Hawaii avec TRIDENT sont présentés. L'analyse des performances indique qu'un compagnon 9,5 magnitudes plus faible qu'une étoile est détectable (6$\sigma$) à 0,5$^{\prime \prime}$ de celle-ci. Les observations suggèrent que les aberrations non communes des trois chemins optiques de la caméra limitent la soustraction des fonctions d'étalement. Un nouveau concept de caméra basé sur un détecteur multibande utilisant une matrice de microlentilles et de microfiltres est décrit pour résoudre le problème des aberrations non communes. Une autre technique d'imagerie, l'imagerie différentielle angulaire, est présentée pour obtenir des fonctions d'étalement de référence dans chaque chemin optique à même les observations. Cette technique s'emploie avec des télescopes spatiaux ou des télescopes terrestres à monture altitude/azimut et consiste à tourner le télescope ou à attendre que le champ tourne suffisamment pour soustraire la fonction d'étalement et conserver le signal des compagnons. Finalement, la précision d'algorithmes de recentrage, de changement d'échelle et de rotation d'images basés sur la transformée de Fourier est étudiée pour montrer qu'ils sont suffisamment précis, en théorie, pour la détection d'un compagnon plus de 10$^{9}$ fois plus faible qu'une étoile.

Mots cl\'{e}s fran\c{c}ais: Astronomie, imagerie, infrarouge, optique adaptative, \'{e}toiles proches, syst\`{e}mes plan\'{e}taires, naines brunes/exoplan\`{e}tes

Pour information: marois@astro.umontreal.ca
\pagebreak[4]
\chapter*{Summary}
\addcontentsline{toc}{chapter}{Summary}
\thispagestyle{myheadings}

The technique of simultaneous spectral differential imaging (SSDI) is used to image directly brown dwarfs and exoplanets around nearby stars. SSDI consists of acquiring a number of images simultaneously at different adjacent narrowband wavelengths and combining them to attenuate atmospheric speckles and noise associated with quasi-static and static wavefront errors to reach photon noise limited PSF subtraction. Wavelengths are selected across the 1.6 $\mu $m methane bandhead so that methanated companions are brighter in one wavelength than at others. This choice ensures that a significant fraction of the methanated companion flux is retained at all separations after combining the images. An analytical PSF model is first presented with numerical simulations to estimate the PSF noise attenuation performance. A three-wavelength infrared camera (TRIDENT) implementing the SSDI technique is then described. Results from a survey of 35 nearby stars carried out with TRIDENT at the Canada-France-Hawaii telescope are presented. Performance estimates show that a companion 9.5 magnitudes fainter than a star is detectable (6$\sigma$) at 0.5$^{\prime \prime}$ separation. An analysis of the observations suggests that non-common path aberrations between TRIDENT optical channels are the limiting factor preventing further PSF noise attenuation. A new camera concept using a multi-wavelength detector featuring a microlens array combined with micro-filters is presented to overcome the non-common path aberration problem. Another imaging technique, differential angular imaging, is also discussed to obtain a reference PSF in each optical channel while observing a target. This technique can be used with space- and ground-based altitude/azimuth telescopes. It consists of rotating the telescope or waiting for sufficient field rotation to subtract the PSF and to preserve the companion flux. Finally, the accuracy of FFT-based image shifting, scaling and rotating algorithms is studied to show that a companion 10$^{9}$ times fainter than a star can theoretically be detected by those algorithms.

Keywords: Astronomy, imaging, infrared, adaptive optic, nearby stars, planetary systems, brown dwarfs/exoplanets

For information: marois@astro.umontreal.ca




\end{document}